\documentclass[aps,twocolumn,a4paper,10pt]{revtex4-1}
\usepackage{mathrsfs}
\usepackage{graphicx}
\usepackage{latexsym}
\usepackage{amsmath}
\usepackage{amssymb}
\usepackage{textcomp}
\usepackage{amsbsy}
\usepackage{graphics}
\usepackage{epstopdf}
\usepackage{color}

\begin{document}
\title{Scalaron tunneling and the fate of antisymmetric tensor fields in $F(R)$ gravity}
\author{Tanmoy Paul}
\email{pul.tnmy9@gmail.com}
\author{Soumitra SenGupta}
\email{tpssg@iacs.res.in}
\affiliation{School of Physical Sciences,\\
Indian Association for the Cultivation of Science,\\
2A $\&$ 2B Raja S.C. Mullick Road,\\
Kolkata - 700 032, India.\\}

\begin{abstract}
The work provides a possible explanation of a well motivated question - why the present universe is practically free from any noticeable 
footmarks of higher rank antisymmetric tensor fields, despite having the signatures of scalar, vector, fermion as well as symmetric 
rank 2 tensor field in the form of gravity ? The explanation proposed here originates from the
higher curvature degrees of freedom present in a $F(R)$  gravity model.
In such a model, we show that the scalar degree of freedom (also known as scalaron) 
associated with the higher curvature term may undergo a quantum tunneling which in turn suppresses
the couplings of antisymmetric massless tensor fields with various Standard Model fields.
\end{abstract}

\maketitle

\section{Introduction}
An astonishing feature of our present universe is that it carries observable footprints of scalar, fermion and 
vector degrees of freedom along with spin 2 massless graviton, 
while there is no noticeable observable effect of any massless 
antisymmetric tensor modes. In general, the antisymmetric tensor fields transform as p-forms and appear as a massless 
representation of Lorentz group. Besides such Lorentz group representation, the antisymmetric fields also introduce themselves as closed string mode 
in String theory \cite{duff}. In this context, the second rank antisymmetric tensor field, generally known as Kalb-Ramond (KR) 
field \cite{kalb} has been studied in a wide domain  of  research from cosmology/Astrophysics to elementary particle 
phenomenology \cite{ssg_KR1,ssg_KR2,ssg_KR3,suppressionF(R)1,suppressionF(R)2,
tp_KR1,tp_KR2,biswarup_KR,sumanta_KR,Paul:2020bdy,Aashish:2019ykb,Aashish:2019zsy}. 
Furthermore it has been shown that KR field ($B_{\mu\nu}$) can also 
act as a possible source of spacetime torsion where the torsion is identified with rank-3 antisymmetric field strength tensor $H_{\mu\nu\rho}$ 
having a relation 
with $B_{\mu\nu}$ as $H_{\mu\nu\rho}=\partial_{[\mu}B_{\nu\rho]}$ \cite{ssg_KR1}. 
The Kalb-Ramond coupling strength to other 
matter fields turns out to  be same as gravity-matter coupling i.e $1/M_p$ where $M_p$ is the reduced Planck mass defined by
$M_p = \frac{1}{\sqrt{8\pi G}}$ with $G$ being the Newton's constant. But no experimental evidences support this argument, in particular, 
all Cosmological/Astrophysical as well as particle phenomenology experiments 
so far have produced only negative results in terms of detecting any signature of antisymmetric tensor fields. 
This implies there must be some underlying mechanism which resulted into suppression of the coupling of  antisymmetric tensor fields to a negligible value
at the present energy scale of our universe. However these couplings 
may be large in early phase of the universe, in particular during inflation where the energy scale (i.e Hubble parameter) ranges from 
$10^{14}-10^{16}\mathrm{GeV}$. This allows us to introduce the question : why the 
effects of the Kalb-Ramond field or any higher rank antisymmetric tensor field are invisibly small in comparison to that of  the gravitational force ? 
One of the possible explanations of this question comes through the extra dimensional braneworld model 
in a warped background, where, due to the exponential warping of the bulk spacetime, the couplings of antisymmetric fields 
are found to be heavily suppressed on the TeV brane \cite{1,2,3,4}.\\
However the present paper attacks this problem in an new way, in particular, we show 
that the suppression of the Kalb-Ramond and also of the other higher rank antisymmetric tensor fields can 
be argued even in four dimensional spacetime without bringing in any extra dimension , with the help of higher 
curvature F(R) gravity in 4 dimensional spacetime. The underlying diffeomorphism invariance allows to generalize the Einstein-Hilbert action by 
introducing higher curvature terms in the gravitational action. Similar  higher curvature terms also appear naturally  in string-inspired models 
due to quantum 
corrections. Among various higher curvature models proposed so far, 
\cite{Nojiri:2010wj,Nojiri:2017ncd,Capozziello:2011et,Artymowski:2014gea,
Nojiri:2003ft,Odintsov:2019mlf,Johnson:2019vwi,Pinto:2018rfg,Odintsov:2019evb,Nojiri:2019riz,Nojiri:2019fft,Lobo:2008sg,
Gorbunov:2010bn,Nojiri:2019lqw,Elizalde:2019tee,Li:2007xn,Odintsov:2020nwm,Odintsov:2020iui,Appleby:2007vb,Elizalde:2010ts,Cognola:2007zu}, 
Gauss-Bonnet (GB) gravity \cite{Li:2007jm,Odintsov:2018nch,Carter:2005fu,Nojiri:2019dwl,Elizalde:2010jx,Makarenko:2016jsy,
delaCruzDombriz:2011wn,Chakraborty:2018scm,
Kanti:2015pda,Kanti:2015dra,Elizalde:2020zcb,Odintsov:2018zhw,Saridakis:2017rdo,Cognola:2006eg}
or more generally Lanczos-Lovelock  gravity \cite{lanczos,lovelock}  drew lot of attention due to their ability to address various issues 
in different sectors of gravitational physics. In 
general the higher curvature terms lead to instabilities such as  Ostragradsky instability, 
instability of cosmological scalar and tensor primordial perturbations or the matter instability in the astrophysical sector. However the Gauss-Bonnet 
model, a special case of the Lanczos-Lovelock gravity, is free from the Ostragradsky instability 
due to the special choice of  coefficients in the Riemann tensor, Ricci tensor and curvature scalar respectively. On other hand, unlike 
the GB theory, the F(R) model contains higher curvature terms consisting only the Ricci scalar which may lead to Ostragradsky instability. 
Such models are often plagued with the presence of ghost terms.
However there exist a class of  F(R) models which are free from such instability. A F(R) model is regarded as a viable one if it is free of the ghosts. 
In general, a F(R) theory can be mapped to a scalar-tensor theory by a Legendre transformation of the metric, where 
the scalar field potential in the scalar-tensor theory depends on the form of F(R) 
\cite{barrow,marino,bahamonde,Das:2017htt,Banerjee:2017lxi,catena,ssg1,ssg2}. 
The issue of the instability of the original F(R) model can 
be transformed to the corresponding dual scalar-tensor theory via the kinetic and potential energy of the scalar field 
( See section II  for a detailed discussion ). Actually 
the scalar potential comes with a stable minimum and the kinetic term will appear with  proper  non-phantom like signature only 
if the original $F(R)$ theory is free from  ghosts.

\section{The model}\label{sec_model}
We begin with a second rank antisymmetric Kalb-Ramond (KR) field in $F(R)$ gravity theory, while the other higher rank antisymmetric 
fields will be considered at a later stage. The
KR field  couples other matter fields, in particular with fermion and U(1)
gauge field and  the observable signatures of the KR field
in our universe is determined by these interaction terms. Thus the action is given by,
\begin{eqnarray}
 S = \int d^4x \sqrt{-g}&\bigg[&\frac{F(R)}{2\kappa^2} - \frac{1}{4}H_{\mu\nu\rho}H^{\mu\nu\rho} 
 + \bar{\Psi}i\gamma^{\mu}\partial_{\mu}\Psi\nonumber\\ 
 &-&\frac{1}{4}F_{\mu\nu}F^{\mu\nu} - \frac{1}{M_p}\bar{\Psi}\gamma^{\mu}\sigma^{\nu\rho}H_{\mu\nu\rho}\Psi\nonumber\\ 
 &-&\frac{1}{M_p}A^{[\mu}F^{\nu\rho]}H_{\mu\nu\rho}\bigg]
 \label{action1}
\end{eqnarray}
where $H_{\mu\nu\rho} = \partial_{[\mu}B_{\nu\rho]}$ is the field strength tensor of the Kalb-Ramond field $B_{\nu\rho}$ and 
$\kappa^2 = 8\pi G = \frac{1}{M_p^2}$ (with $G$ being the Newton's constant and $M_p \sim 10^{18}\mathrm{GeV}$ is the reduced Planck mass). 
Apart from the $F(R)$ term in the action (\ref{action1}), the second, third and fourth terms denote the kinetic part of the KR field $B_{\mu\nu}$, 
spin $\frac{1}{2}$ fermions $\Psi$ and $U(1)$ gauge 
field $A_{\mu}$ respectively, moreover the interaction between $B_{\mu\nu}-\Psi$ and $B_{\mu\nu}-A_{\mu}$ are represented by the last two terms of the 
action. The interaction between KR and U(1) electromagnetic field originates from Chern-Simons term which is incorporated 
to make the theory free from gauge anomaly.\\
The above action, by introducing an auxiliary field $A(x)$, can be equivalently expressed as,
\begin{eqnarray}
 S&=& \int d^4x \sqrt{-g}\bigg[\frac{1}{2\kappa^2}\bigg(F'(A)(R-A)+F(A)\bigg)\nonumber\\ 
 &-&\frac{1}{4}H_{\mu\nu\rho}H^{\mu\nu\rho} + \bar{\Psi}i\gamma^{\mu}\partial_{\mu}\Psi 
 - \frac{1}{4}F_{\mu\nu}F^{\mu\nu}\nonumber\\ 
 &-&\frac{1}{M_p}\bar{\Psi}\gamma^{\mu}\sigma^{\nu\rho}H_{\mu\nu\rho}\Psi - \frac{1}{M_p}A^{[\mu}F^{\nu\rho]}H_{\mu\nu\rho}\bigg]
 \label{action2}
\end{eqnarray}
The auxiliary field equation leads to the solution $A=R$ which, along with the action (\ref{action2}), immediately reproduces the 
initial action given in Eq.(\ref{action1}). This confirms the equivalence between the two actions. At this stage, we consider a 
transformation of the metric as
\begin{equation}
 g_{\mu\nu}(x) \longrightarrow \widetilde{g}_{\mu\nu}(x) = e^{-\sqrt{\frac{2}{3}}\kappa \xi(x)}g_{\mu\nu}(x)
 \label{conformal}
\end{equation}
$\mu, \nu$ run form 0 to 3. $\xi(x)$ is conformal factor and is related to the auxiliary 
field as $e^{-\sqrt{\frac{2}{3}}\kappa \xi} = F'(A)$. The gamma matrices satisfy the algebra $\{\gamma^{\mu},\gamma^{\nu}\} = 2g^{\mu\nu}$ and 
$\sigma^{\nu\rho}$ is the commutator of gamma matrices, in particular $\sigma^{\nu\rho} = \frac{i}{2}[\gamma^{\nu},\gamma^{\rho}]$ \cite{stephani}. Thus due 
to Eq.(\ref{conformal}), the transformations of $\gamma^{\mu}$ and $\sigma^{\nu\rho}$ are given by,
\begin{eqnarray}
 \gamma^{\mu} \longrightarrow \widetilde{\gamma}^{\mu} = e^{\frac{1}{2}\sqrt{\frac{2}{3}}\kappa \xi} \gamma^{\mu}
 \nonumber
 \end{eqnarray}
 and
 \begin{eqnarray}
 \sigma^{\nu\rho} \longrightarrow \widetilde{\sigma}^{\nu\rho} = e^{\sqrt{\frac{2}{3}}\kappa \xi} \sigma^{\nu\rho}
 \nonumber
\end{eqnarray}
respectively. Moreover, if $R$ and $\tilde{R}$ are the Ricci scalars corresponding to the metrics $g_{\mu\nu}$ and 
$\tilde{g}_{\mu\nu}$ respectively, then these are related as,
\begin{eqnarray}
 R =e^{-\Big{(}\sqrt{\frac{2}{3}}\kappa\xi\Big{)}}\Big{(}\tilde{R} - \kappa^2\tilde{g}^{\mu\nu}\partial_{\mu}\xi\partial_{\nu}\xi
 + 2\kappa\sqrt{\frac{3}{2}}\tilde{\Box}\xi \Big{)}\, .
 \nonumber
\end{eqnarray}
where $\tilde{\Box}$ represents the d'Alembertian operator in terms of the metric tensor $\tilde{g}_{\mu\nu}$. Due to the above 
relation between $R$ and $\tilde{R}$, the action (\ref{action2}) can be written as,
\begin{eqnarray}
 S&=&\int d^4x \sqrt{-\tilde{g}}\nonumber\\
 &\bigg[&\frac{1}{2\kappa^2} e^{[\sqrt{\frac{2}{3}}\kappa\xi]}F'(A)\bigg(\tilde{R}
 - \kappa^2\tilde{g}^{\mu\nu}\partial_{\mu}\xi\partial_{\nu}\xi
 + 2\kappa \sqrt{\frac{3}{2}}\tilde{\Box}\xi\bigg)\nonumber\\
 &-&\frac{1}{2\kappa^2}e^{2[\sqrt{\frac{2}{3}}\kappa\xi]}\bigg(AF'(A) - F(A)\bigg)\nonumber\\ 
 &+&e^{2[\sqrt{\frac{2}{3}}\kappa\xi]} \bigg(-\frac{1}{4}H_{\mu\nu\rho}H^{\mu\nu\rho} + \bar{\Psi}i\gamma^{\mu}\partial_{\mu}\Psi 
 - \frac{1}{4}F_{\mu\nu}F^{\mu\nu}\nonumber\\ 
 &-&\frac{1}{M_p}\bar{\Psi}\gamma^{\mu}\sigma^{\nu\rho}H_{\mu\nu\rho}\Psi - \frac{1}{M_p}A^{[\mu}F^{\nu\rho]}H_{\mu\nu\rho}\bigg)\bigg]
 \label{action intermediate}
\end{eqnarray}

\begin{figure}[!h]
\begin{center}
 \centering
 \includegraphics[width=3.0in,height=2.5in]{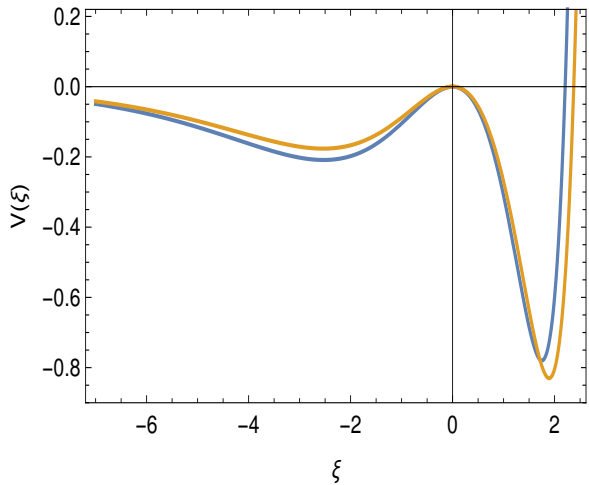}
 \caption{$Blue~Curve$: $V(\xi)$ (along y-axis) vs $\xi$ (along x-axis) for $\frac{\sqrt{\beta}}{\alpha}=0.53$. 
 $Yellow~Curve$: $V(\xi)$ (along y-axis) vs $\xi$ (along x-axis) for $\frac{\sqrt{\beta}}{\alpha}=0.525$.}
 \label{plot potential}
\end{center}
\end{figure}

Considering $F'(A) > 0$ and using the relation between $\xi$ and $F'(A)$, one obtains the following scalar-tensor action,

\begin{eqnarray}
 S&=&\int d^4x \sqrt{-\tilde{g}}\bigg[\frac{\widetilde{R}}{2\kappa^2} + \frac{1}{2}\tilde{g}^{\mu\nu}\partial_{\mu}\xi \partial_{\nu}\xi 
 - \bigg(\frac{AF'(A) - F(A)}{2\kappa^2F'(A)^2}\bigg)\nonumber\\
 &-&\frac{1}{4}e^{-\sqrt{\frac{2}{3}}\kappa \xi} H_{\mu\nu\rho}H_{\alpha\beta\delta} 
 \tilde{g}^{\mu\alpha}\tilde{g}^{\nu\beta}\tilde{g}^{\rho\delta} 
 - \frac{1}{4}F_{\mu\nu}F_{\alpha\beta} \tilde{g}^{\mu\alpha}\tilde{g}^{\nu\beta}\nonumber\\
 &+&e^{\sqrt{\frac{2}{3}}\kappa \xi} {\Psi}^{+}\tilde{\gamma}^{0}i\tilde{\gamma}^{\mu}\partial_{\mu}\Psi 
 - \frac{1}{M_p} {\Psi}^{+}\tilde{\gamma}^{0}\tilde{\gamma}^{\mu}\tilde{\sigma}^{\nu\rho}H_{\mu\nu\rho}\Psi\nonumber\\
 &-&\frac{1}{M_p} e^{-\sqrt{\frac{2}{3}}\kappa \xi} A_{[\alpha}F_{\beta\delta]}H_{\mu\nu\rho} 
 \tilde{g}^{\mu\alpha}\tilde{g}^{\nu\beta}\tilde{g}^{\rho\delta}\bigg]
 \label{action3}
\end{eqnarray}
where $\widetilde{R}$ is the Ricci scalar formed by $\widetilde{g}_{\mu\nu}$. The field $\xi(x)$ acts as a scalar field 
endowed with a potential $\frac{AF'(A) - F(A)}{2\kappa^2F'(A)^2}$ ($= V(A(\xi))$, say). Thereby it can be argued that the higher curvature 
degree(s) of freedom manifests itself through a scalar field degree of freedom $\xi(x)$ with a potential $V(\xi)$ which in turn 
depends on the form of $F(R)$. Further it is also important to note that for $F'(R) < 0$, the kinetic term 
of the scalar field $\xi$, as well as the Ricci scalar $R$ in the 
above action come with wrong sign, which indicates the existence 
of a ghost field. To avoid this, the 
derivative of the functional form of $F(R)$ gravity, namely $F'(R)$ must be chosen to be greater than zero.\\
In view of action (\ref{action3}), here it deserves mentioning that the appearance of the scalaron field $\xi(x)$ makes the fermion field 
$\Psi$ and the KR field $B_{\mu\nu}$ non-canonical, while at the same time the electromagnetic field remains canonical. Thus the fields are redefined 
to make their respective kinetic terms canonical, as follows:
\begin{eqnarray}
 B_{\mu\nu} \longrightarrow \widetilde{B}_{\mu\nu} = e^{-\frac{1}{2}\sqrt{\frac{2}{3}}\kappa \xi} B_{\mu\nu}
 \label{ft_1}
 \end{eqnarray}
 \begin{eqnarray}
 \Psi \longrightarrow \widetilde{\Psi} = e^{\frac{1}{2}\sqrt{\frac{2}{3}}\kappa \xi} \Psi
 \label{ft_2}
 \end{eqnarray}
 and
 \begin{eqnarray}
 A_{\mu} \longrightarrow \widetilde{A}_{\mu} = A_{\mu}
 \label{ft_3}
\end{eqnarray}
In terms of these redefined fields, the scalar-tensor action becomes canonical and is given by,
\begin{eqnarray}
 S&=&\int d^4x \sqrt{-\tilde{g}}\bigg[\frac{\widetilde{R}}{2\kappa^2} + \frac{1}{2}\tilde{g}^{\mu\nu}\partial_{\mu}\xi \partial_{\nu}\xi 
 - V(\xi)\nonumber\\
 &-&\frac{1}{4} \tilde{H}_{\mu\nu\rho}\tilde{H}^{\mu\nu\rho}  
 - \frac{1}{4}\tilde{F}_{\mu\nu}\tilde{F}^{\mu\nu} + \bar{\tilde{\Psi}}i\tilde{\gamma}^{\mu}\partial_{\mu}\tilde{\Psi}\nonumber\\ 
 &-&\frac{1}{M_p} e^{(-\frac{1}{2}\sqrt{\frac{2}{3}}\kappa \xi)} 
 \bar{\tilde{\Psi}}\tilde{\gamma}^{\mu}\tilde{\sigma}^{\nu\rho}\tilde{H}_{\mu\nu\rho}\tilde{\Psi}\nonumber\\
 &-&\frac{1}{M_p} e^{(-\frac{1}{2}\sqrt{\frac{2}{3}}\kappa \xi)} \tilde{A}_{[\alpha}\tilde{F}_{\beta\delta]}\tilde{H}_{\mu\nu\rho}\nonumber\\
 &+&terms~proportional~to~\partial_{\mu}\xi\bigg]
 \label{action4}
\end{eqnarray}
The above scalar-tensor canonical action shows that the interaction terms (between $\tilde{B}_{\mu\nu}$ and $\tilde{\Psi}$, $\tilde{A_{\mu}}$) 
present in the corresponding Lagrangian carry an exponential factor 
$e^{(-\frac{1}{2}\sqrt{\frac{2}{3}}\kappa \xi)}$ over $1/M_p$ by which gravity couples with matter fields. Our main goal is to investigate whether 
such exponential factor (in front of the interaction terms) suppresses the coupling strengths of KR field with various matter fields 
which in turn may explain the invisibility of the Kalb-Ramond field in our present universe. For this purpose, we need a certain form of the scalar 
field potential $V(\xi)$ and recall, $V(\xi)$ in turn depends on the form of $F(R)$. The stability 
of $V(\xi)$ follows from the following two conditions on $F(R)$:

\begin{equation}
 \bigg[2F(R) - RF'(R)\bigg]_{\langle R\rangle} = 0
 \label{stability condition1}
\end{equation}
and
\begin{equation}
 \bigg[\frac{F'(R)}{F''(R)} - R\bigg]_{\langle R\rangle} > 0
 \label{stability condition2}
\end{equation}
In order to 
achieve an explicit expression of a stable scalar potential, we choose the form of $F(R)$ as a polynomial of Ricci scalar of order 3 as,

\begin{eqnarray}
 F(R) = R - \alpha R^2 + \beta R^3
 \label{form}
\end{eqnarray}
where $\alpha$ and $\beta$ are the model parameters with mass dimensions [-2] and [-4] respectively. 
For this specific choice of $F(R)$, the potential $V(\xi)$ becomes,

\begin{eqnarray}
 V(\xi)&= \frac{\alpha^3}{54\kappa^2\beta^2} e^{2\sqrt{\frac{2}{3}}\kappa\xi}
 \bigg[\sqrt{1 - \frac{3\beta}{\alpha^2} + \frac{3\beta}{\alpha^2} e^{-\sqrt{\frac{2}{3}}\kappa\xi}} - 1\bigg]^2\nonumber\\
 &\bigg[1 - 2\sqrt{1 - \frac{3\beta}{\alpha^2} + \frac{3\beta}{\alpha^2} e^{-\sqrt{\frac{2}{3}}\kappa\xi}}\bigg]
 \label{potential}
\end{eqnarray}
This potential has two minima at,

\begin{eqnarray}
 \langle\xi\rangle_\mathrm{-}&=&\sqrt{\frac{3}{2\kappa^2}} \ln{\bigg[\frac{1}{4+2\alpha/\sqrt{\beta}}\bigg]} < 0
 \label{vev1}
 \end{eqnarray}
 \begin{eqnarray}
 \langle\xi\rangle_\mathrm{+}&=&\sqrt{\frac{3}{2\kappa^2}} \ln{\bigg[\frac{1}{4-2\alpha/\sqrt{\beta}}\bigg]} > 0
 \label{vev2}
\end{eqnarray}
and a maximum at

\begin{eqnarray}
 \langle\xi\rangle_\mathrm{max} = 0
 \label{maxima}
\end{eqnarray}
(which can also be shown from the expression $[2F(R) - RF'(R)]_{\langle R\rangle} = 0$, see Eqns.(\ref{stability condition1}), (\ref{stability condition2})), 
as long as the parameters satisfy the condition as $\frac{1}{2} < \frac{\sqrt{\beta}}{\alpha} < \frac{1}{\sqrt{3}}$, this indicates 
that $\alpha$ and $\beta$ are greater 
than zero. It may also be mentioned that 
the same conditions on $\alpha$ and $\beta$ make the potential $V(\xi)$ - a real valued function. Eqn.(\ref{potential}) clearly indicates that 
asymptotically $V(\xi)$ tends to $-e^{\frac{1}{2}\sqrt{\frac{2}{3}}\kappa\xi} \rightarrow 0^{-}$ as $\xi$ goes to $-\infty$ and 
diverges for $\xi \rightarrow \infty$. Moreover 
$V(\xi)$ has two zero at $\xi_1 = 0$ and $\xi_2 = \sqrt{\frac{3}{2\kappa^2}}\ln{\bigg[\frac{1}{1 - \frac{\alpha^2}{4\beta}}\bigg]}$ 
$\big($ $\xi_2$ is greater than $\langle\xi\rangle_\mathrm{+}$ as $\frac{\alpha}{2\sqrt{\beta}} < 1$, see above$\big)$ respectively. With these 
features, we present a plot of $V(\xi)$ vs. $\xi$ in Fig.[\ref{plot potential}] for two different values of $\frac{\sqrt{\beta}}{\alpha}$, 
in particular for $\frac{\sqrt{\beta}}{\alpha} = 0.53$ and $\frac{\sqrt{\beta}}{\alpha} = 0.525$ respectively.\\

As mentioned earlier and also evident from Fig.[\ref{plot potential}], $V(\xi)$ has two minima at $\langle\xi\rangle_\mathrm{-}$ and at 
$\langle\xi\rangle_\mathrm{+}$. Thereby from stability 
criteria if the scalar field $\xi$ is at $\langle\xi\rangle_\mathrm{-}$, then it is in a local minimum and in order to have 
a lower energy configuration, the scalar field may tunnel from $\xi = \langle\xi\rangle_\mathrm{-}$ to $\xi = \langle\xi\rangle_\mathrm{+}$ 
which is a global minimum. Such tunneling 
may happen due to quantum fluctuations.\\
Below, we discuss how this tunneling may lead to the invisibility of second rank antisymmetric Kalb-Ramond field in our present universe.\\

\section{Suppression of antisymmetric tensor fields due to tunneling}
\begin{figure}[!h]
\begin{center}
 \centering
 \includegraphics[width=3.5in,height=2.5in]{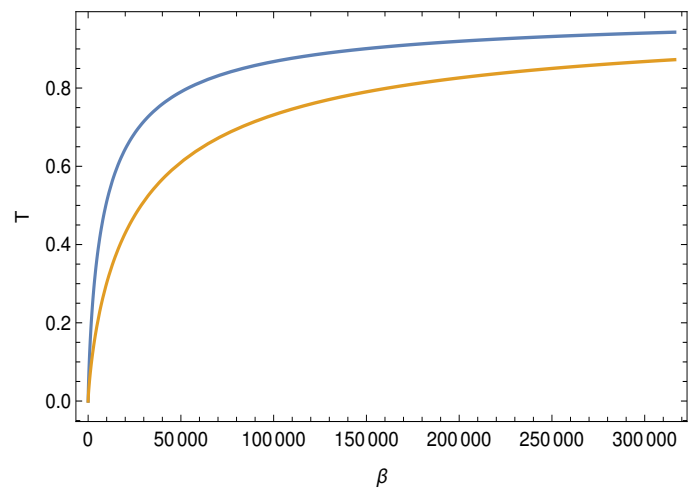}
 \caption{$Blue~Curve$: $T$ (along y-axis) vs $\beta$ (along x-axis) for $\frac{\sqrt{\beta}}{\alpha} = 0.53$. 
 $Yellow~Curve$: $T$ (along y-axis) vs $\beta$ (along x-axis) for $\frac{\sqrt{\beta}}{\alpha} = 0.525$.}
 \label{plot probability1}
\end{center}
\end{figure}

In this case, as pointed out earlier, the scalar field tunnels from $\langle\xi\rangle_\mathrm{-}$ to $\langle\xi\rangle_\mathrm{+}$ 
in order to attain a lower energy configuration.  We will 
see later that such tunneling effect is intimately connected to the explanation of the invisibility of the signature of  KR field in our universe. 
To calculate the tunneling probability from $\langle\xi\rangle_\mathrm{-}$ to $\langle\xi\rangle_\mathrm{+}$, 
the scalar potential is approximately 
taken as a rectangle barrier having width ($w$) $=\kappa^2[\langle\xi\rangle_\mathrm{+} - \langle\xi\rangle_\mathrm{-}]$ and height 
($h$) $=V(\langle\xi\rangle_\mathrm{+})$ respectively. 
For such a potential barrier, the standard quantum mechanical tunneling probability ($T$) is given by \cite{coleman1,gervais,Darme:2019ubo},

\begin{eqnarray}
 \frac{1}{T} = 1 + \bigg|\frac{V(\langle\xi\rangle_\mathrm{+})}{V(\langle\xi\rangle_\mathrm{-})}
 \bigg| \sinh^2\bigg[w\sqrt{\frac{2m_{\xi}\Delta V}{M^3}}\bigg]
 \label{tunneling1}
\end{eqnarray}
where $m_{\xi}^2 = V''(\langle\xi\rangle_\mathrm{-})$ is the mass of the scalaron field and obtained as,
\begin{eqnarray}
 m_{\xi}^2 = \frac{3}{2\sqrt{\beta}(4+2\alpha/\sqrt{\beta})(6+2\alpha/\sqrt{\beta})}
 \label{mass}
\end{eqnarray}
Moreover the width of the barrier and $\Delta V = V(\langle\xi\rangle_\mathrm{-}) - V(\langle\xi\rangle_\mathrm{+})$ takes the following form,
\begin{eqnarray}
 w = \sqrt{\frac{3}{2\kappa^2}} \ln{\bigg[\frac{4 + 2\alpha/\sqrt{\beta}}{4 - 2\alpha/\sqrt{\beta}}\bigg]}
 \label{width}
\end{eqnarray}
and
\begin{eqnarray}
 \Delta V = \frac{\alpha(12+\alpha^2/\beta)}{4\kappa^2\beta(4-\alpha^2/\beta)^2}
 \label{delta V}
\end{eqnarray}
respectively. Plugging these expressions into Eqn.(\ref{tunneling1}) leads to an  explicit form of the tunneling probability as,
\begin{eqnarray}
 \frac{1}{T}&= 1 + \bigg(\frac{4 + 2\alpha/\sqrt{\beta}}{4 - 2\alpha/\sqrt{\beta}}\bigg)^3
 \sinh^2\bigg[\sqrt{6}\ln{\bigg(\frac{4 + 2\alpha/\sqrt{\beta}}{4 - 2\alpha/\sqrt{\beta}}\bigg)}\times\nonumber\\
 &\sqrt{\frac{\sqrt{12}\big(12 + \alpha^2/\beta\big)~\alpha/\sqrt{\beta}}
 {M^3\beta^{3/4}\big(4+2\alpha/\sqrt{\beta}\big)^{5/2}\big(4-2\alpha/\sqrt{\beta}\big)^{2}\big(6+2\alpha/\sqrt{\beta}\big)^{1/2}}}\bigg]
 \label{tunneling2}
\end{eqnarray}
Eqn.(\ref{tunneling2}) clearly indicates that $T$ depends on $\alpha/\sqrt{\beta}$ and $\beta$. Recall that the ratio $\frac{\sqrt{\beta}}{\alpha}$ is 
constrained by 
\begin{eqnarray}
\frac{1}{2} < \frac{\sqrt{\beta}}{\alpha} < \frac{1}{\sqrt{3}}
\label{constraint}
\end{eqnarray}
Keeping this in mind, we give the plot of $T$ vs. $\beta$ (see Fig.[\ref{plot probability1}]) for two different values of 
$\frac{\sqrt{\beta}}{\alpha}$, in particular for $\frac{\sqrt{\beta}}{\alpha} = 0.53,0.525$ respectively.\\

\begin{figure}[!h]
\begin{center}
 \centering
 \includegraphics[width=3.5in,height=2.5in]{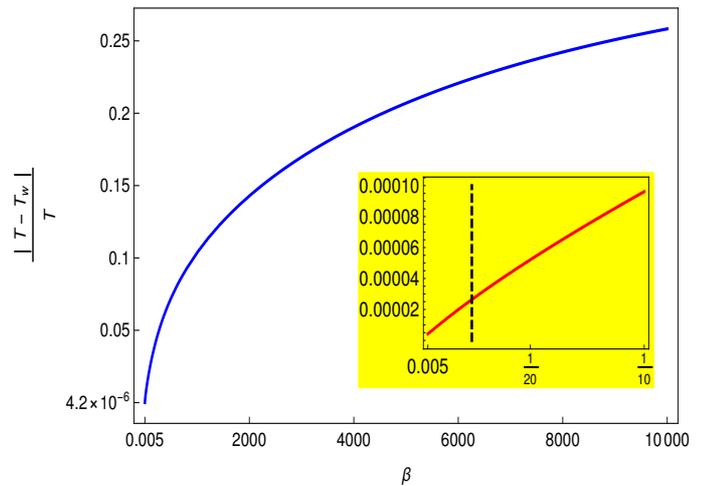}
 \caption{$\frac{|T-T_{w}|}{T}$ (along y-axis) vs $\beta$ (along x-axis, in Planckian unit) for a wide range of $\beta$. 
 The inset-graph depicts the behaviour of $\frac{|T-T_{w}|}{T}$ for $0.005 \leq \beta \leq 0.1$ (in Planckian unit) by the solid red curve. 
 The time scale for the scalaron tunneling is of the order $\sim 10^{17}$ sec. (i.e the present age of the universe) 
 which actually occurs near $\beta = 2.43\times10^{-2}$. This value of $\beta$ is represented by the vertical dashed line in the inset graph, which 
 shows that for $\beta = 2.43\times10^{-2}$, the fractional difference $\frac{|T-T_{w}|}{T}$ becomes $\sim 10^{-5}$.}
 \label{plot probability_comparison}
\end{center}
\end{figure}

Fig.[\ref{plot probability1}] reveals that the tunneling probability increases with the value of the higher curvature parameter $\beta$ 
and asymptotically reaches to unity at large $\beta$. However this is 
expected, because with increasing value of $\beta$, the height of the potential barrier ($\propto \frac{1}{\sqrt{\beta}}$) decreases and 
as a consequence, $T$ increases. Moreover, $T$ goes to zero as $\beta$ tends to zero, because for $\beta \rightarrow 0$, 
the height of the potential barrier goes to infinite and as a result, $T = 0$. 
Furthermore the time scale for tunneling is related to the inverse 
of the tunneling probability $T^{-1}$ which decreases with the increasing value of the higher curvature parameter $\beta$ or $\alpha$. As mentioned earlier, 
our motivation is to show that the scalaron tunneling results into the suppression of the couplings of KR field with other matter fields 
in the present universe. We propose that the time scale for the tunneling from $\langle\xi\rangle_\mathrm{-}$ to $\langle\xi\rangle_\mathrm{+}$ is of 
the order of the present age of the universe i.e 
\begin{eqnarray}
 \frac{\kappa}{T} \simeq 10^{17} sec. \simeq 10^{41}\mathrm{GeV}^{-1}
 \label{estimation1}
\end{eqnarray}
where the conversion $1$sec. $= 10^{24}\mathrm{GeV}^{-1}$ may be useful. Recall, the parameters $\alpha$ and $\beta$ are constrained by 
$\frac{\sqrt{\beta}}{\alpha} > \frac{1}{2}$; thus by considering $\frac{\sqrt{\beta}}{\alpha} = 0.53$ (the Fig.[\ref{plot probability1}] is also 
plotted for $\frac{\sqrt{\beta}}{\alpha} = 0.53$) we estimate the parameter $\beta$ for which Eqn.(\ref{estimation1}) is satisfied. As a result, 
$\beta = 2.43\times10^{-2}\kappa^4$ makes $T^{-1} \simeq 10^{60}$ leading to the time scale of the scalaron tunneling of 
the order of the present age of our universe. We will use this fact later. At this stage it deserves mentioning that we have  calculated the tunneling 
probability by approximating the scalaron potential as a rectangular barrier having finite width and height. Thus it becomes 
important to investigate whether this approximation is a viable one in the present context and for this purpose we try to determine the tunneling 
probability by relaxing the ``rectangular barrier'' approximation. In particular taking the actual potential 
we use the WKB method which leads to the scalaron tunneling 
probability from $\langle\xi\rangle_\mathrm{-}$ to $\langle\xi\rangle_\mathrm{+}$ as,
\begin{eqnarray}
 T_w \simeq e^{-2\gamma}
 \label{WKB 1}
\end{eqnarray}
where the subscript 'w' symbolizes for WKB method and $\gamma$ takes the following form
\begin{eqnarray}
 \gamma&=&\int_{\langle\xi\rangle_\mathrm{-}}^{\xi_0} d\xi~\kappa^2\sqrt{\frac{2m_{\xi}}{M^3}\big(V(\xi) - V(\langle\xi\rangle_\mathrm{-})\big)}\nonumber\\ 
 &+&\int_{\xi_0}^{\langle\xi\rangle_\mathrm{+}} d\xi~\kappa^2\sqrt{\frac{2m_{\xi}}{M^3}\big(V(\langle\xi\rangle_\mathrm{-}) - V(\xi)\big)}
 \label{WKB 2}
\end{eqnarray}
with $\xi_0$ being the turning point given by $V(\xi) = V(\langle\xi\rangle_\mathrm{-})$. It may be observed that in-between 
$\langle\xi\rangle_\mathrm{-}$ and $\xi_0$, the energy 
of the scalaron field is less than that of the corresponding potential, while the situation gets reversed for $\xi = [\xi_0,\langle\xi\rangle_\mathrm{+}]$. 
Using Eq.(\ref{WKB 1}) we estimate the tunneling probability obtained from the WKB method 
and examine whether such estimation matches with the estimate based on ``rectangular barrier'' approximation. The form of the scalaron potential 
(see Fig.[\ref{plot potential}]) indicates that the main contribution of $\gamma$ comes from the region $\xi = [\xi_0,\langle\xi\rangle_\mathrm{+}]$ where 
$V(\xi)$ can be taken as $V(\langle\xi\rangle_\mathrm{+})$ and therefore 
$\gamma$ can be expressed as 
$\gamma = \kappa^2\big(\langle\xi\rangle_\mathrm{+} - \xi_0\big)
\sqrt{\frac{2m_{\xi}}{M^3}\big(V(\langle\xi\rangle_\mathrm{-}) - V(\langle\xi\rangle_\mathrm{+})\big)}$. Moreover 
$m_{\xi}^2$ and $\Delta V$ are determined in Eqs.(\ref{mass}) and (\ref{delta V}) respectively. These expressions yield the explicit form of the WKB 
tunneling probability from Eqn.(\ref{WKB 1}) as,
\begin{eqnarray}
 &\ln{\big(T_w\big)} = -2\sqrt{6}\ln{\bigg(\frac{1}{4 - 2\alpha/\sqrt{\beta}}\bigg)}&\nonumber\\
 &\times\sqrt{\frac{\sqrt{12}\big(12 + \alpha^2/\beta\big)~\alpha/\sqrt{\beta}}
 {M^3\beta^{3/4}\big(4+2\alpha/\sqrt{\beta}\big)^{5/2}\big(4-2\alpha/\sqrt{\beta}\big)^{2}\big(6+2\alpha/\sqrt{\beta}\big)^{1/2}}}&\nonumber\\
 \label{WKB 3}
\end{eqnarray}
In order to investigate the viability of rectangular barrier approximation, we present a plot depicting the fractional 
difference of the tunneling probabilities (with respect to $\beta$) based on the rectangular barrier approximation and on the WKB method respectively, 
i.e $\frac{|T-T_{w}|}{T}$ vs. $\beta$, see Fig.[\ref{plot probability_comparison}] where $\beta$ is taken in the Planckian unit. 
The figure demonstrates that the fractional difference between $T$ and $T_w$ is negligible 
for small values of $\beta$ which however becomes significant at large $\beta$. 
To get a better view of what is happening near low values of $\beta$, 
we give a zoomed-in version by the inset-graph in Fig.[\ref{plot probability_comparison}] showing 
the behaviour of $\frac{|T-T_{w}|}{T}$ for $0.005 \leq \beta \leq 0.1$ (in Planckian unit) by the solid red curve, while the dashed vertical line of the 
inset graph represents $\beta = 2.43\times10^{-2}$ which is of interest regime of $\beta$ in the present context. 
Here, our motivation is to provide a possible explanation for the 
suppression of the KR field on the present universe due to  the scalaron tunneling, for which the time scale of the tunneling 
should be of the order of the age of the present universe i.e $\frac{\kappa}{T} \sim 10^{17}$ sec. or equivalently 
$T \simeq 10^{-60}$ which actually occurs for $\beta = 2.43\times10^{-2}$ (in Planckian unit), as mentioned earlier. 
Moreover the inset-graph of Fig.[\ref{plot probability_comparison}] clearly shows that for $\beta = 2.43\times10^{-2}$ (in Planckian unit), 
the fractional difference $\frac{|T-T_{w}|}{T}$ is of the order of $10^{-5}$. These arguments indicate 
that the rectangular barrier approximation is a viable consideration, however up to an error of $10^{-5}$, 
in the present context where the time scale for the tunneling is of the order $\sim 10^{17}$ sec.\\
Before the tunneling when the scalar field is frozen at its vev $\langle\xi\rangle_\mathrm{-}$, it becomes non-dynamical and thus the action 
in eqn.(\ref{action4}) turns out to be,
\begin{eqnarray}
 S&=&\int d^4x \sqrt{-\tilde{g}}\bigg[\frac{\widetilde{R}}{2\kappa^2} - \frac{1}{4} \tilde{H}_{\mu\nu\rho}\tilde{H}^{\mu\nu\rho}  
 - \frac{1}{4}\tilde{F}_{\mu\nu}\tilde{F}^{\mu\nu}\nonumber\\ 
 &+&\bar{\tilde{\Psi}}i\tilde{\gamma}^{\mu}\partial_{\mu}\tilde{\Psi} - \frac{1}{M_p} e^{(-\frac{1}{2}\sqrt{\frac{2}{3}}\kappa \langle\xi\rangle_\mathrm{-})} 
 \bar{\tilde{\Psi}}\tilde{\gamma}^{\mu}\tilde{\sigma}^{\nu\rho}\tilde{H}_{\mu\nu\rho}\tilde{\Psi}\nonumber\\
 &-&\frac{1}{M_p} e^{(-\frac{1}{2}\sqrt{\frac{2}{3}}\kappa \langle\xi\rangle_\mathrm{-})} \tilde{A}^{[\mu}\tilde{F}^{\nu\rho]}\tilde{H}_{\mu\nu\rho}\bigg]
 \label{action5}
\end{eqnarray}
The last two terms in the above expression of action give the coupling of KR field to fermion, 
 $U(1)$ gauge field before the tunneling and are given by 
 \begin{eqnarray}
  \lambda^{(b)}_{KR-fer}&=&\frac{1}{M_p} e^{(-\frac{1}{2}\sqrt{\frac{2}{3}}\kappa \langle\xi\rangle_\mathrm{-})}\nonumber\\
  &=&\frac{1}{M_p} \sqrt{4 + 2\alpha/\sqrt{\beta}}
  \label{couplingbefore1}
 \end{eqnarray}
 and
 \begin{eqnarray}
  \lambda^{(b)}_{KR-U(1)}&=&\frac{1}{M_p} e^{(-\frac{1}{2}\sqrt{\frac{2}{3}}\kappa \langle\xi\rangle_\mathrm{-})}\nonumber\\
  &=&\frac{1}{M_p} \sqrt{4 + 2\alpha/\sqrt{\beta}}
  \label{couplingbefore2}
 \end{eqnarray}
 respectively. Similarly after the tunneling when the scalar field acquires the vev $\langle\xi\rangle_\mathrm{+}$, the couplings of KR field to fermion 
 and $U(1)$ gauge field are given by
 \begin{eqnarray}
  \lambda^{(a)}_{KR-fer}&=&\frac{1}{M_p} e^{(-\frac{1}{2}\sqrt{\frac{2}{3}}\kappa \langle\xi\rangle_\mathrm{+})}\nonumber\\
  &=&\frac{1}{M_p} \sqrt{4 - 2\alpha/\sqrt{\beta}}
  \label{couplingafter1}
 \end{eqnarray}
 and
 \begin{eqnarray}
  \lambda^{(a)}_{KR-U(1)}&=&\frac{1}{M_p} e^{(-\frac{1}{2}\sqrt{\frac{2}{3}}\kappa \langle\xi\rangle_\mathrm{+})}\nonumber\\
  &=&\frac{1}{M_p} \sqrt{4 - 2\alpha/\sqrt{\beta}}
  \label{couplingafter2}
 \end{eqnarray}
  where we use the explicit expression of $\langle\xi\rangle_\mathrm{+}$. As the ratio $\frac{\alpha}{\sqrt{\beta}}$ is constrained to be less 
  than $2$ (see Eqn.(\ref{constraint})), there is no worry about the realness of $\lambda^{(a)}$(s). However it may be observed 
  that $\lambda^{(a)}$(s) are smaller than $\lambda^{(b)}$(s), which 
  clearly argues that the couplings of KR field (with matter fields) get reduced due to the tunneling of the scalar field. 
  More explicitly, Eqns.(\ref{couplingbefore1}) to 
  (\ref{couplingafter2}) demonstrate that before the tunneling, interactions of KR field with various matter fields are of same order 
  as usual gravity-matter coupling strength $1/M_p$, 
 while after the tunneling, the couplings of the KR field get suppressed by the factor $\sqrt{4 - 2\alpha/\sqrt{\beta}}$ 
 and the suppression increases as the ratio of $\alpha/\sqrt{\beta}$ approaches more closer to the value $2$. Moreover as we showed earlier, 
 the time scale of the scalaron tunneling or equivalently the decaying time scale of the KR field is of the order 
 of the present age of the universe. These arguments indicate 
 that the effect of KR field may be dominant in the early universe but as the universe evolves, the scalar field makes 
 a tunneling from $\xi = \langle\xi\rangle_\mathrm{-}$ to $\xi = \langle\xi\rangle_\mathrm{+}$ (with a probability $T$, see eqn.(\ref{tunneling2})) 
 which induces a suppression on the interaction strengths of KR field. This fact is in agreement with \cite{suppressionF(R)1, suppressionF(R)2,tp_KR1} 
 and may explain the negligible footprint of Kalb-Ramond field 
 on our present universe.\\
 
 Having discussing about the KR field, now we move forward to the third rank antisymmetric tensor field 
 $X_{\alpha\beta\rho}$ for which the corresponding 'field 
 strength tensor' is given by $Y_{\alpha\beta\rho\delta}$ ($= \partial_{[\alpha}X_{\beta\rho\delta]}$). The action for $X_{\alpha\beta\rho}$ 
 in four dimension is expressed as, 
 \begin{eqnarray}
  S[X] = \int d^4x \sqrt{-g} Y_{\alpha\beta\rho\delta}Y^{\alpha\beta\rho\delta}
  \nonumber
 \end{eqnarray}
The same procedure, as discussed above for KR field, leads to the coupling of rank 3 antisymmetric field $X$ to matter (before and after the tunneling) as,
 \begin{eqnarray}
  \Omega^{(b)}_{X-fer} = \frac{1}{M_p} \sqrt{4 + 2\alpha/\sqrt{\beta}}\nonumber\\
  \Omega^{(b)}_{X-U(1)} = \frac{1}{M_p} \big[4 + 2\alpha/\sqrt{\beta}\big]
  \label{coupling before higher1}
 \end{eqnarray}
 and
 \begin{eqnarray}
  \Omega^{(a)}_{X-fer} = \frac{1}{M_p} \sqrt{4 - 2\alpha/\sqrt{\beta}}\nonumber\\
  \Omega^{(a)}_{X-U(1)} = \frac{1}{M_p} \big[4 - 2\alpha/\sqrt{\beta}\big]
  \label{coupling4}
 \end{eqnarray}
 where $\Omega^{(b)}_{X-fer}$ and $\Omega^{(a)}_{X-fer}$ denote the coupling between $X-$fermion before and after the tunneling of the scalaron 
 respectively (same type of symbols are used for $X-U(1)$ gauge field). 
 Like the case of KR field, $\Omega^{(a)}_{X-fer}$ and $\Omega^{(a)}_{X-U(1)}$ get suppressed in comparison to that before the tunneling. 
 However $\lambda^{(a)}_{KR-fer}$ and $\Omega^{(a)}_{X-fer}$ carry the same suppression factor while the comparison of 
 $\lambda^{(a)}_{KR-U(1)}$ and $\Omega^{(a)}_{X-U(1)}$ makes it clear that the interaction with electromagnetic 
 field become progressively smaller with the increasing rank of the tensor field. Therefore  the suppression of antisymmetric tensor fields due to the  tunneling of the scale field increases with the increase in the rank of the tensor field.\\
 These are demonstrated in Table[1].
 \begin{table}[!h]
  \centering
  \resizebox{\columnwidth}{1.3 cm}{%
  \begin{tabular}{|c| c| c| c| c| c|}
   \hline 
   $\frac{2\alpha}{\sqrt{\beta}}$ & Coupling before tunneling  & $\lambda^{(a)}_{KR-fer}$ & $\lambda^{(a)}_{KR-U(1)}$ & $\Omega^{(a)}_{X-fer}$ 
   & $\Omega^{(a)}_{X-U(1)}$\\
   \hline
   $4-10^{-2}$ & $\sim \frac{1}{M_p}$ & $10^{-1}/M_p$ & $10^{-1}/M_p$ & $10^{-1}/M_p$ & $10^{-2}/M_p$ \\
   $4-10^{-4}$ & $\sim \frac{1}{M_p}$ & $10^{-2}/M_p$ & $10^{-2}/M_p$ & $10^{-2}/M_p$ & $10^{-4}/M_p$\\
   $4-10^{-8}$ & $\sim \frac{1}{M_p}$ & $10^{-4}/M_p$ & $10^{-4}/M_p$ & $10^{-4}/M_p$ & $10^{-8}/M_p$\\
    \hline
  \end{tabular}%
  }
  \caption{Couplings of antisymmetric tensor fields for a wide range of $\alpha/\sqrt{\beta}$, before and after the tunneling of the scalaron}
  \label{Table-1}
 \end{table}
 
 It may be mentioned that the Standard Model gauge couplings remain invariant under conformal transformation of the metric 
 while the Yukawa coupling parameters can be suitably redefined.\\
 Till now we considered  a certain form of $F(R) = R - \alpha R^2 + \beta R^3$, as shown in eqn.(\ref{form}). Therefore a natural question will be whether our findings are confined only  within this particular form of $F(R)$ or there exist a class of F(R) for which the strong 
 suppression of the torsion coupling to Standard Model fields would occur naturally. We discuss it in the following section.\\
 
 \section{Class of F(R) models consistent with strong suppression of the torsion coupling}
 At this stage, it becomes clear that the F(R) models which correspond to a scalaron potential having two minima and one maximum can suppress 
 the coupling between Kalb-Ramond and Standard Model fields at the present universe and thus fulfill our purpose. Recall that the extrema of 
 scalaron potential can be determined from $\bigg[2F(R) - RF'(R)\bigg]_{\langle R \rangle} = 0$ and the maximum or minimum depend on the condition whether 
 the expression $\bigg[\frac{F'(R)}{F''(R)} - R\bigg]_{\langle R \rangle}$ is greater or less than zero 
 ( see Eqns.(\ref{stability condition1}) and (\ref{stability condition2}) ). 
 Therefore in order to have two minima and one maximum ( i.e three extrema ) of the 
 scalaron potential, $F(R)$ must contain a term with cubic or higher order of Ricci curvature otherwise the expression 
 $\big[2F(R) - RF'(R)\big]_{\langle R \rangle} = 0$ does not produce three extrema.\\
 With these informations in hand, we consider a general class of polynomial 
 $F(R) = R - \alpha R^2 + \beta R^m$ model with $m \geq 3$ and investigate whether such F(R) models can lead to negligible footprints of KR field 
 in our present universe. The scalaron potential corresponds to this F(R) is given by,
 \begin{eqnarray}
  V(\xi) = \frac{1}{2\kappa^2} \frac{\bigg((m-1)\beta R^m - \alpha R^2\bigg)}{\bigg(1 - 2\alpha R + m\beta R^{m-1}\bigg)^2}
  \label{general case_potential}
 \end{eqnarray}
where $R = R(\xi)$ can be obtained by the following relation, 
 \begin{eqnarray}
  e^{-\sqrt{\frac{2}{3}}\kappa\xi} = 1 - 2\alpha R + m\beta R^{m-1}
  \label{general case_relation}
 \end{eqnarray}
 In order to invert Eqn.(\ref{general case_relation}) to get the functional behaviour of $R = R(\xi)$, one needs a certain value of the 
 parameter $m$. In a later part of this section, we will present the explicit form of the scalaron potential by considering a particular value of $m$. 
 However before moving to this part, let us discuss how far can we go with the generalized version i.e without considering any particular value of $m$ and 
 for this purpose we will address the cases for odd values of $m$ and for even values of $m$ separately in the following two subsections.\\
 
 \subsection{Case-I: Odd values of m}
\begin{figure}[!h]
\begin{center}
 \centering
 \includegraphics[width=3.2in,height=2.5in]{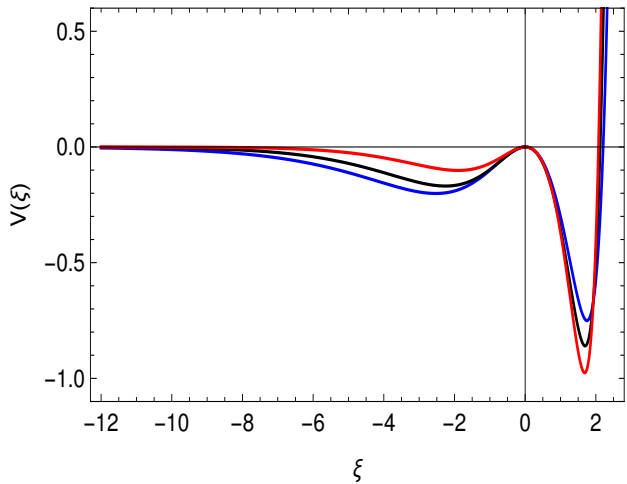}
 \caption{$Blue~Curve$: $V(\xi)$ (along y-axis) vs $\xi$ (along x-axis) for $m = 3$ and $\frac{\beta}{\alpha^2}=0.28$. 
 $Black~Curve$: $V(\xi)$ (along y-axis) vs $\xi$ (along x-axis) for $m = 5$ and $\frac{\beta}{\alpha^4}=0.25$. 
 $Red~Curve$: $V(\xi)$ (along y-axis) vs $\xi$ (along x-axis) for $m = 7$ and $\frac{\beta}{\alpha^6}=0.15$. For 
 $m = 3$ the parameter $\frac{\beta}{\alpha^2}$ is dimensionless while for $m = 5$ and $m = 7$, $\frac{\beta}{\alpha^4}$ and 
 $\frac{\beta}{\alpha^6}$ respectively are dimensionless.}
 \label{plot potential_particular}
\end{center}
\end{figure}

 For odd values of $m$, the potential in Eqn.(\ref{general case_potential}) has 
 two minima ($\langle\xi\rangle_\mathrm{-}$ and $\langle\xi\rangle_\mathrm{+}$) and one maximum ($\langle\xi\rangle_\mathrm{max}$ 
 between the two minima) at
 \begin{eqnarray}
  \langle\xi\rangle_\mathrm{-} = \sqrt{\frac{3}{2\kappa^2}} \ln{\bigg[\frac{1}{\frac{2m-2}{m-2} + 2\alpha \big(\frac{1}{\beta(m-2)}\big)^{\frac{1}{m-1}}}\bigg]}
  \label{general_vev1}
 \end{eqnarray}
\begin{eqnarray}
  \langle\xi\rangle_\mathrm{+} = \sqrt{\frac{3}{2\kappa^2}} \ln{\bigg[\frac{1}{\frac{2m-2}{m-2} - 2\alpha \big(\frac{1}{\beta(m-2)}\big)^{\frac{1}{m-1}}}\bigg]}
  \label{general_vev2}
 \end{eqnarray}
and
\begin{eqnarray}
 \langle\xi\rangle_\mathrm{max} = 0
 \label{general_maxima}
\end{eqnarray}
respectively with the parameters satisfying the condition $\alpha \big(\frac{1}{\beta(m-2)}\big)^{\frac{1}{m-1}} < (m-1)/(m-2)$. To obtain the 
above extrema of $V(\xi)$ in terms of $\xi$, we use the relation $R = R(\xi)$ from Eqn.(\ref{general case_relation}). 
Clearly for $m=3$, the above three expressions resemble with the previous ones as obtained 
in Eqns.(\ref{vev1}), (\ref{vev2}) and (\ref{maxima}) respectively. Moreover the potential (\ref{general case_potential}) has two zero 
at $R_1 = 0$ and at $R_2 = \bigg(\frac{\alpha}{\beta(m-1)}\bigg)^{\frac{1}{(m-2)}}$ respectively. Eqn.(\ref{general case_relation}) immediately 
leads to these zeros in terms of the scalaron field $\xi$ as follows:
 \begin{eqnarray}
 \xi_1&=&0\nonumber\\
  \xi_2&=&\sqrt{\frac{3}{2\kappa^2}} \ln{\bigg[\frac{1}{1 - \alpha\big(\frac{m-2}{m-1}\big)\big(\frac{\alpha}{\beta(m-1)}\big)^{\frac{1}{m-2}}}\bigg]}
  \label{general_zero}
 \end{eqnarray}
 It may be observed that $\xi_1$ coincides with $\langle\xi\rangle_\mathrm{max}$, which in turn indicates that 
 $V(\langle\xi\rangle_\mathrm{max}) = 0$ i.e the potential 
 vanishes at the point of its maximum (recall from Fig.[\ref{plot potential}] that the potential for $m = 3$ has two zeros and one of them matches with 
 its maximum). Furthermore, the $V(\xi)$ of eqn.(\ref{general case_potential}) at the respective minimum acquires the following expressions:
 \begin{eqnarray}
  V(\langle\xi\rangle_\mathrm{-})&=&-\frac{1}{4\kappa^2}\bigg(\frac{1}{\beta(m-2)}\bigg)^{\frac{1}{(m-1)}}\nonumber\\
  &\bigg\{&\frac{\bigg[\frac{2m-2}{m-2} - 2\alpha \big(\frac{1}{\beta(m-2)}\big)^{\frac{1}{m-1}}\bigg]}
  {\bigg[\frac{2m-2}{m-2} + 2\alpha \big(\frac{1}{\beta(m-2)}\big)^{\frac{1}{m-1}}\bigg]^2}\bigg\}
  \label{general_value2}
 \end{eqnarray}
 and
\begin{eqnarray}
 V(\langle\xi\rangle_\mathrm{+})&=&-\frac{1}{4\kappa^2}\bigg(\frac{1}{\beta(m-2)}\bigg)^{\frac{1}{(m-1)}}\nonumber\\
  &\bigg\{&\frac{\bigg[\frac{2m-2}{m-2} + 2\alpha \big(\frac{1}{\beta(m-2)}\big)^{\frac{1}{m-1}}\bigg]}
  {\bigg[\frac{2m-2}{m-2} - 2\alpha \big(\frac{1}{\beta(m-2)}\big)^{\frac{1}{m-1}}\bigg]^2}\bigg\}
  \label{general_value3}
\end{eqnarray}
Due to the aforementioned constraint $\alpha \big(\frac{1}{\beta(m-2)}\big)^{\frac{1}{m-1}} < (m-1)/(m-2)$, $V(\langle\xi\rangle_\mathrm{-})$ and 
$V(\langle\xi\rangle_\mathrm{+})$ become negative and also $\big|V(\langle\xi\rangle_\mathrm{+})\big| > \big|V(\langle\xi\rangle_\mathrm{-})\big|$. 
Thus the scalaron potential corresponds 
to $F(R) = R - \alpha R^2 + \beta R^m$ allows a tunneling of the scalaron field from $\langle\xi\rangle_\mathrm{-}$ to $\langle\xi\rangle_\mathrm{+}$ 
in order to attain 
a lower energy configuration. Such tunneling effect leads to a negligible footprint of the Kalb-Ramond field in our present universe. 
Once again, the scalar potential is approximately 
taken as a rectangle barrier having width ($w$) $=\kappa^2[\langle\xi\rangle_\mathrm{+} - \langle\xi\rangle_\mathrm{-}]$, height 
($h$) $=V(\langle\xi\rangle_\mathrm{+})$ 
and consequently the tunneling probability ($T$) is obtained as,
\begin{eqnarray}
 \frac{1}{T} = 1 + \bigg|\frac{V(\langle\xi\rangle_\mathrm{+})}{V(\langle\xi\rangle_\mathrm{-})}\bigg| 
 \sinh^2\bigg[w\sqrt{\frac{2m_{\xi}\Delta V}{M^3}}\bigg]
 \label{general_tunneling1}
\end{eqnarray}
where $m_{\xi}$ and $\Delta V$ are given by,
\begin{eqnarray}
 m_{\xi}^2&=&V''(\langle\xi\rangle_\mathrm{-})\nonumber\\
 &=&\frac{3(m-1)}{2\big(\beta(m-2)\big)^{\frac{1}{(m-1)}}\bigg[m\big(\frac{m-1}{m-2}\big) + 2\alpha \big(\frac{1}{\beta(m-2)}\big)^{\frac{1}{m-1}}\bigg]}\nonumber\\
 &\times&\frac{1}{\bigg[\frac{2m-2}{m-2} + 2\alpha \big(\frac{1}{\beta(m-2)}\big)^{\frac{1}{m-1}}\bigg]}
 \label{general_mass}
\end{eqnarray}
and
\begin{eqnarray}
 \Delta V&=&\frac{\alpha}{\kappa^2}\bigg(\frac{1}{\beta(m-2)}\bigg)^{\frac{2}{(m-1)}}\nonumber\\ 
 &\bigg\{&\frac{\bigg[3\big(\frac{2m-2}{m-2}\big)^2 + 4\alpha^2\big(\frac{1}{\beta(m-2)}\big)^{\frac{2}{m-1}}\bigg]}
 {\bigg[\big(\frac{2m-2}{m-2}\big)^2 - 4\alpha^2\big(\frac{1}{\beta(m-2)}\big)^{\frac{2}{m-1}}\bigg]^2}\bigg\}
 \label{general_delta}
\end{eqnarray}
respectively. Plugging back the above expressions into Eqn.(\ref{general_tunneling1}), one gets the final form of the tunneling probability as follows,
\begin{eqnarray}
 \frac{1}{T}&=1 + \bigg(\frac{p + q}{p - q}\bigg)^3 
 \sinh^2\bigg[\frac{\sqrt{3}}{2}\ln{\big(\frac{p+q}{p-q}\big)}\times\nonumber\\
 &\sqrt{\frac{\sqrt{6(m-1)}(3p^2 + q^2)q}{M^3\big(\beta(m-2)\big)^{\frac{3}{2m-2}}
 (p + q)^{5/2}(p - q)^2(\frac{m}{2}p + q)^{1/2}}}~\bigg]
 \label{general_tunneling2}
\end{eqnarray}
where $p = \frac{2m - 2}{m - 2}$ and $q = 2\alpha\bigg(\frac{1}{\beta(m-2)}\bigg)^{\frac{1}{(m-1)}}$. 
With Eqns.(\ref{general_vev1}) and (\ref{general_vev2}) in hand, 
we follow the similar procedure as discussed earlier to yield the couplings of KR field before and after tunneling as  
 \begin{eqnarray}
  \lambda^{(b)}_{KR-fer}&=&\lambda^{(b)}_{KR-U(1)}\nonumber\\
  &=&\frac{1}{M_p} \bigg[\frac{2m-2}{m-2} + 2\alpha \big(\frac{1}{\beta(m-2)}\big)^{1/(m-1)}\bigg]^{1/2}\nonumber\\
  \label{couplingbefore_revised}
 \end{eqnarray}
 Similarly after the tunneling when the scalar field acquires the vev $\langle\xi\rangle_\mathrm{+}$, the couplings of KR field to fermion 
 and $U(1)$ gauge field are given by
 \begin{eqnarray}
  \lambda^{(a)}_{KR-fer}&=&\lambda^{(a)}_{KR-U(1)}\nonumber\\
  &=&\frac{1}{M_p} \bigg[\frac{2m-2}{m-2} - 2\alpha \big(\frac{1}{\beta(m-2)}\big)^{1/(m-1)}\bigg]^{1/2}\nonumber\\
  \label{couplingafter_revised}
 \end{eqnarray}
 Clearly $\lambda^{(a)} < \lambda^{(b)}$ which entails that 
 the interactions of KR field with various matter fields get suppressed due to the  effect 
 of scalaron tunneling. This may explain why the present universe is practically free from observable signatures 
 of KR field. Thus the class of $F(R) = R - \alpha R^2 + \beta R^m$ with odd values of $m$ fit well and serve the purpose in 
 the present context.\\
 Having described the general version for odd values of $m$, now we present explicit form of a scalaron potential by considering a certain value 
 of $m$, in particular for $m = 5$ (the $m = 3$ case has already  been discussed in Sec.[\ref{sec_model}]). Solving 
 Eqn.(\ref{general case_relation}) with $m = 5$, one gets the functional behaviour of $R = R(\xi)$ as follows,
 \begin{eqnarray}
  R(\xi) = \frac{\sqrt{f(\xi)}}{2\sqrt{15}}\bigg[1 + \sqrt{1 + \frac{12\sqrt{15}\alpha}{\beta}}\frac{1}{f(\xi)^{3/2}}\bigg]
  \label{relation_particular case}
 \end{eqnarray}
where $f(\xi)$ has the following expression,
\begin{eqnarray}
f(\xi)&=&\frac{30^{1/3}}{\beta}\bigg[9\alpha^2\beta + \sqrt{3}\beta\sqrt{27\alpha^4 - 80\beta\big(1 - e^{-\sqrt{\frac{2}{3}}\kappa\xi}\big)^3}\bigg]^{1/3}\nonumber\\
&+&\frac{2\times30^{2/3}\big(1 - e^{-\sqrt{\frac{2}{3}}\kappa\xi}\big)}
{\bigg[9\alpha^2\beta + \sqrt{3}\beta\sqrt{27\alpha^4 - 80\beta\big(1 - e^{-\sqrt{\frac{2}{3}}\kappa\xi}\big)^3}\bigg]^{1/3}}
 \label{f}
\end{eqnarray}
The function $f(\xi)$ is real values for $\frac{\beta}{\alpha^4} < \frac{27}{80}$. Plugging the above expression of $R(\xi)$ into 
Eqn.(\ref{general case_potential}) yields the scalaron potential for $m=5$ as,
\begin{eqnarray}
 &V(\xi)&= \frac{1}{8\kappa^2}e^{2\sqrt{\frac{2}{3}}\kappa\xi}\nonumber\\
 &\bigg[&\frac{\beta}{2}\bigg\{\sqrt{\frac{f(\xi)}{15}}\bigg(1 + \sqrt{1 + \frac{12\sqrt{15}\alpha}{\beta}}\frac{1}{f(\xi)^{3/2}}\bigg)\bigg\}^5\nonumber\\
 &-&\alpha\bigg\{\sqrt{\frac{f(\xi)}{15}}\bigg(1 + \sqrt{1 + \frac{12\sqrt{15}\alpha}{\beta}}\frac{1}{f(\xi)^{3/2}}\bigg)\bigg\}^2\bigg]
 \label{potential_particular case}
\end{eqnarray}
with $f(\xi)$ being given in Eqn.(\ref{f}). 
Here we give the plot of the scalaron potential obtained in Eq.(\ref{potential_particular case}) in Fig.[\ref{plot potential_particular}]. 
Moreover to compare the scalaron potential for various values of $m$, we also present the plot of $V(\xi)$ for $m = 3$ and $m = 7$ as well in the same 
Fig.[\ref{plot potential_particular}].\\

Fig.[\ref{plot potential_particular}] depicts that the scalaron potential corresponding to $m = 5$ has two minima (say at $\langle\xi\rangle_{-}$ 
and at $\langle\xi\rangle_{+}$ 
where $\big|V(\langle\xi\rangle_{-})\big| < \big|V(\langle\xi\rangle_{+})\big|$) and one maximum between the two minima. 
Thus the potential allows a quantum tunneling of the 
scalaron field from $\langle\xi\rangle_{-}$ to $\langle\xi\rangle_{+}$ in order to have lower energy configuration, 
in effect of which the couplings of the KR field with other matter fields get suppressed.

\subsection{Case-II: Even values of $m$}
 The $F(R) = R - \alpha R^2 + \beta R^m$ with even values of $m$ corresponds to such a scalar field potential 
 which has two real extrema, as an example, for $m = 4$, the scalaron potential has the extrema at
 \begin{eqnarray}
  \langle\xi\rangle = 0~~~~~~~~~,~~~~~~\langle\xi\rangle = \sqrt{\frac{3}{2\kappa^2}} \ln{\bigg[\frac{1}{3 - \frac{2\alpha}{(2\beta)^{1/3}}}\bigg]}
 \nonumber
\end{eqnarray}
respectively. Moreover for $m=6$, the potential acquires the extrema at $\langle\xi\rangle = 0$ and 
$\langle\xi\rangle = \sqrt{\frac{3}{2\kappa^2}} \ln{\bigg[\frac{1}{\frac{5}{2} - \frac{2\alpha}{(4\beta)^{1/5}}}\bigg]}$. Recall, in order 
to explain the suppression of the KR field in our present universe, the scalaron potential should have three extrema, in particular, two minima 
and one maximum in between the two minima. However as we have examined that the scalaron potential corresponding to 
$F(R) = R - \alpha R^2 + \beta R^m$ with even values of $m$ has two extrema and thus 
such F(R) model is unable to explain why the KR field couplings to SM matter fields get suppressed 
at our present universe by the scalaron tunneling. The same argument also goes for another class of $F(R) = R + \alpha R^{n}$ ($n \geq 3$) i.e. model without the presence of any quadratic term in $R$.
This also leads to the scalaron potential having two extrema only and thus can not explain the suppression of the KR field due to the scalaron tunneling.\\
 
 \section{Equivalence between Kalb-Ramond and a pseudoscalar field in the present context}
 Before concluding, it may be mentioned that the Kalb-Ramond field can be equivalently written as derivative of a pseudoscalar field ($\Phi(x)$, 
 known as axion field) by the following relation:
 \begin{eqnarray}
  H_{\mu\nu\alpha} = \varepsilon_{\mu\nu\alpha\beta}\partial^{\beta}\Phi
  \label{revised1}
 \end{eqnarray}
where $\varepsilon_{\mu\nu\alpha\beta} = \sqrt{-g}[\mu\nu\alpha\beta]$ is the four dimensional Levi-Civita tensor. Therefore 
$\varepsilon_{\mu\nu\alpha\beta}$ transforms in effect of transformation of the metric 
($g_{\mu\nu}(x) \longrightarrow \widetilde{g}_{\mu\nu}(x) = e^{-\sqrt{\frac{2}{3}}\kappa \xi(x)}g_{\mu\nu}(x)$, see eqn.(\ref{conformal})) as ,
\begin{eqnarray}
 \varepsilon_{\mu\nu\alpha\beta} \longrightarrow \tilde{\varepsilon}_{\mu\nu\alpha\beta} 
 = e^{-2\sqrt{\frac{2}{3}}\kappa \xi(x)} \varepsilon_{\mu\nu\alpha\beta}
 \label{revised2}
\end{eqnarray}

Using Eqn.(\ref{revised1}), we 
express the kinetic term of KR field in terms of axion field as follows:
\begin{eqnarray}
 S_{KR}&=&\int d^4x \sqrt{-g} \bigg[-\frac{1}{12}H_{\mu\nu\alpha}H^{\mu\nu\alpha}\bigg]\nonumber\\
 &=&\int d^4x \sqrt{-g} \bigg[-\frac{1}{12}\varepsilon_{\mu\nu\alpha\beta}\varepsilon^{\mu\nu\alpha\sigma}\partial^{\beta}\Phi\partial_{\sigma}\Phi\bigg]
 \nonumber\\
 &=&\int d^4x \sqrt{-g} \bigg[-\frac{1}{2}g^{\mu\nu}\partial_{\mu}\Phi\partial_{\nu}\Phi\bigg]\nonumber\\
 &=&S[\Phi]
 \label{revised3}
\end{eqnarray}
where we use the multiplicative property of Levi-Civita tensor as 
 $\varepsilon^{\mu\nu\alpha\beta}\varepsilon_{\mu\nu\alpha\sigma} = 3! \delta^{\beta}_{\sigma}$. Thus the kinetic term of the axion field is free from 
 any Levi-Civita tensor, as expected, while the interaction terms of axion with SM mater fields contain $\varepsilon_{\mu\nu\alpha\beta}$. 
 Therefore it is important to study whether the procedure of scalaron tunneling can yield the same suppression factor
 (i.e $e^{-\frac{1}{2}\sqrt{\frac{2}{3}}\kappa \xi(x)}$, see Eqn.(\ref{action5})) on the $axion$ couplings with various SM fields.\\
 In terms of the axion field the action of our model takes the following form:
 \begin{eqnarray}
 S = \int d^4x \sqrt{-g}&\bigg[&\frac{F(R)}{2\kappa^2} - \frac{1}{2}g^{\mu\nu}\partial_{\mu}\Phi\partial_{\nu}\Phi 
 + \bar{\Psi}i\gamma^{\mu}\partial_{\mu}\Psi\nonumber\\ 
 &-&\frac{1}{4}F_{\mu\nu}F^{\mu\nu} - \frac{1}{M_p}\varepsilon_{\mu\nu\rho\beta}\bar{\Psi}\gamma^{\mu}\sigma^{\nu\rho}\partial^{\beta}\Phi\Psi\nonumber\\ 
 &-&\frac{1}{M_p}\varepsilon_{\mu\nu\rho\beta}A^{[\mu}F^{\nu\rho]}\partial^{\beta}\Phi\bigg]
 \label{action_revised1}
 \end{eqnarray}
 
 As shown earlier , the model can be recast to a scalar-tensor theory by a Legendre transformation of the metric and in effect of such 
 transformation the kinetic terms of the fields become non-canonical. The following redefinitions make the respective kinetic terms canonical,
 \begin{eqnarray}
 \Phi \longrightarrow \widetilde{\Phi} = e^{\frac{1}{2}\sqrt{\frac{2}{3}}\kappa \xi} \Phi
 \label{ft_revised1}
 \end{eqnarray}
 \begin{eqnarray}
 \Psi \longrightarrow \widetilde{\Psi} = e^{\frac{1}{2}\sqrt{\frac{2}{3}}\kappa \xi} \Psi
 \label{ft_revised2}
 \end{eqnarray}
 and
 \begin{eqnarray}
 A_{\mu} \longrightarrow \widetilde{A}_{\mu} = A_{\mu}
 \label{ft_revised3}
\end{eqnarray}

Comparing Eqns.(\ref{ft_revised1}) and (\ref{ft_1}), it becomes clear that the transformation 
of $\Phi$ is different in comparison to that of $B_{\mu\nu}$. However it is expected because the 
transformation of the metric also affects the Levi-Civita tensor, as shown in Eqn.(\ref{revised2}). 
In terms of these redefined fields, the scalar-tensor action becomes canonical and is given by,
\begin{eqnarray}
 S&=&\int d^4x \sqrt{-\tilde{g}}\bigg[\frac{\widetilde{R}}{2\kappa^2} + \frac{1}{2}\tilde{g}^{\mu\nu}\partial_{\mu}\xi \partial_{\nu}\xi 
 - V(\xi)\nonumber\\
 &-&\frac{1}{2} \tilde{g}^{\mu\nu}\partial_{\mu}\tilde{\Phi}\partial_{\nu}\tilde{\Phi}  
 - \frac{1}{4}\tilde{F}_{\mu\nu}\tilde{F}^{\mu\nu} + \bar{\tilde{\Psi}}i\tilde{\gamma}^{\mu}\partial_{\mu}\tilde{\Psi}\nonumber\\ 
 &-&\frac{1}{M_p} e^{(-\frac{1}{2}\sqrt{\frac{2}{3}}\kappa \xi)}\tilde{\varepsilon}_{\mu\nu\rho\beta} 
 \bar{\tilde{\Psi}}\tilde{\gamma}^{\mu}\tilde{\sigma}^{\nu\rho}\partial^{\beta}\tilde{\Phi}\tilde{\Psi}\nonumber\\
 &-&\frac{1}{M_p} e^{(-\frac{1}{2}\sqrt{\frac{2}{3}}\kappa \xi)}\tilde{\varepsilon}_{\mu\nu\rho\beta} \tilde{A}_{[\alpha}\tilde{F}_{\beta\delta]}
 \partial^{\beta}\tilde{\Phi}\nonumber\\
 &+&terms~proportional~to~\partial_{\mu}\xi\bigg]
 \label{action_revised4}
\end{eqnarray} 
Comparing Eqns.(\ref{action_revised4}) and (\ref{action4}), we can argue that the interaction terms 
( between $\tilde{\Phi}$ and $\tilde{\Psi}$, $\tilde{A_{\mu}}$ ) in Eqn.(\ref{action_revised4})) carries the same exponential factor 
$e^{(-\frac{1}{2}\sqrt{\frac{2}{3}}\kappa \xi)}$ over the usual gravity-matter coupling $1/M_p$, as obtained previously when we work with 
the three rank antisymmetric tensor field $H_{\mu\nu\rho}$. Now 
following the same procedure as discussed in the earlier sections, we can show that the couplings of the axion field to SM matter fields 
get suppressed, similar to 3 rank tensor field $H_{\mu\nu\rho}$ 
due to the effect of scalaron tunneling at the present stage of our universe. This may explain the equivalence between the KR field tensor 
and the axion field from the perspective of the present paper.\\
At this stage it deserves mentioning that there exist some dark matter models in the literature 
where the axion field is considered to be a possible candidate to solve the mystery 
of dark matter \cite{axion1,axion2,axion3}. However the axion field we discussed in the above section 
is a dual of the antisymmetric Kalb-Ramond field which can also act as a possible source of spacetime torsion \cite{ssg_KR1}, 
unlike to the class of axion discussed in the earlier literature in the context of dark matter \cite{axion1,axion2,axion3}. 
Further in \cite{Morais:2017vlf}, a massive 3-form field has been used to explain the dark energy epoch of the 
present universe. This is different in comparison to our present paper where we deal with the massless Kalb-Ramond field. Recall that the action or in 
particular the interactions of KR field with other matter fields we considered in Eqn.(\ref{action4}) 
are invariant under electromagnetic and KR gauge transformations which is valid only for massless gauge fields.\\
\section{Conclusion}
In conclusion, for every $F(R)$ gravity there exists an intrinsic scalar degree 
of freedom, besides the massless graviton in the dual scalar-tensor theory. The scalar field, also known as scalaron, is endowed with a 
potential term which in turn depends on the form of $F(R)$. 
To fulfill our purpose in the present context, we consider the form of $F(R)$ as a cubic polynomial function of Ricci scalar 
for which the scalar potential 
 ($V(\xi)$) acquires two minima (say at $\langle\xi\rangle_\mathrm{-}$ and $\langle\xi\rangle_\mathrm{+}$) 
 with different energy configurations, in particular 
 $V(\langle\xi\rangle_\mathrm{+}) < V(\langle\xi\rangle_\mathrm{-})$. 
 In such a scenario, we explore the possible effects of the scalaron tunneling on the coupling of antisymmetric tensor fields (rank two or onwards) with 
 various Standard Model 
 fields.\\
From stability criteria, if $\xi$ is at $\langle\xi\rangle_\mathrm{-}$, 
then in order to have a lower energy configuration, the scalar field has a non-zero probability 
to tunnel 
 from $\xi = \langle\xi\rangle_\mathrm{-}$ to $\xi = \langle\xi\rangle_\mathrm{+}$, which occur due to quantum fluctuation. 
 The tunneling probability is explicitly determined 
 (see eqn.(\ref{tunneling2})) and it is found to increase with the larger value of higher curvature parameter. However it turns out that 
 before the tunneling of the scalar field, 
 the interactions of all antisymmetric tensor fields with matter fields are same as $1/M_p$ by which gravity couples with matter fields, 
 while after the tunneling, the couplings get severely suppressed over $\frac{1}{M_p}$. Therefore one can 
 argue that the effects of antisymmetric fields may be dominant in the early universe but as the universe evolved, the scalar field tunnels 
 from $\xi = \langle\xi\rangle_\mathrm{-}$ to $\xi = \langle\xi\rangle_\mathrm{+}$ which in turn induces a 
 suppression on the interaction strengths of antisymmetric fields with standard model fields 
 in comparison to their $1/M_p$ coupling before the tunneling. Furthermore the amount of suppression increases as the rank of the antisymmetric tensor field 
 takes higher value. This may provide a natural 
 explanation why the large scale behaviour of the present universe is free from any perceptible signatures of all antisymmetric tensor fields 
 and solely governed by gravity.\\
 We also show that our findings are not confined only to our considered $F(R) = R - \alpha R^2 + \beta R^3$ model, but also valid for a general 
 class of polynomial $F(R) = R - \alpha R^2 + \beta R^m$ (with odd integer values of $m$) models. Actually any F(R) model(s) that 
 corresponds to a scalaron potential which has two minima and one maximum can yield the suppression to the KR field couplings with various SM 
 matter fields and the class of model $F(R) = R - \alpha R^2 + \beta R^m$ (with odd integer values of $m$) 
 satisfies such conditions.
 
 \section*{Acknowledgments}
 SSG acknowledges the financial support from SERB Research Project No. EMR/2017/001372. 
 TP acknowledges the hospitality by ICE-CSIC/IEEC (Barcelona, Spain), where a part of this work was done.

\end{document}